\documentclass[aps,pra,twocolumn]{revtex4}
\usepackage{bm}
\usepackage{amstext}
\usepackage{amsmath}
\usepackage{epsfig}
\usepackage{rotating}
\usepackage{psfrag}
\usepackage{rotating}
\usepackage{psfrag}

\newlength{\LyXMinipageIndent}
\newcommand{\kbar} {\mathchar'26\mkern-9muk}
\newcommand{\rme}{{\rm e}}

\begin{document}

\title{Experimental investigation of early-time diffusion in
the quantum kicked rotor using a Bose-Einstein condensate}
\author{G.J. Duffy$\dag$}
\author{S. Parkins$\ddag$}
\author{T. M\"uller$\dag$}
\author{M. Sadgrove$\ddag$}
\author{R. Leonhardt$\ddag$}
\author{A. C. Wilson$\dag$}
\affiliation{$\dag$ Department of Physics, University of Otago, P.O. Box 56,
Dunedin, New
Zealand \\
$\ddag$ Department of Physics, University of Auckland, Private Bag 92019,
Auckland, New Zealand }
\date{\today}

\begin{abstract}
We report the experimental observation of resonances in the early-time
momentum diffusion rates for the atom-optical delta-kicked rotor.
In this work a Bose-Einstein condensate provides a source of ultra-cold atoms
with an ultra-narrow initial
momentum distribution, which is then subjected to periodic pulses (or ``kicks'')
using an intense far-detuned optical standing wave. A quantum resonance occurs
when the momentum
eigenstates accumulate the same phase between kicks leading to ballistic
energy growth. Conversely, an anti-resonance is observed when the phase
accumulated from successive kicks cancels and the system returns to its
initial state. Our experimental results are compared with
theoretical predictions.
\end{abstract}

\maketitle

%\pacs{03.75.Lm, 32.80.Qk}

\section{Introduction}

The delta-kicked rotor (DKR) is a nonlinear dynamical system which
exhibits starkly contrasting behaviour in classical and quantum
regimes. For example, the well-known chaotic diffusion exhibited
by the classical DKR is completely suppressed by coherence effects
(dynamical localization \cite{Casati1979,Shepelyansky1981}) in the quantum regime.
The field of quantum chaos (see, e.g., \cite{Reichl1992,Haake1991}),
which brings together the study of classically chaotic systems and
their quantum mechanical analogues, is relatively new, especially
given the maturity of the two parent fields. Indeed, most of the
progress in quantum chaos has been made only during the last quarter
century.

From an experimental point of view, the field of quantum chaos received a major boost in the 1990's with the use of ultra-cold atoms and pulsed
standing-wave laser fields to realize a near-ideal quantum version of
the delta-kicked rotor \cite{Moore1995}.
At the low temperatures achievable using laser cooling, quantum
behavior of the atomic particles becomes manifest, and optical
manipulation of the atoms offers unprecedented control over the forces
they experience.
Furthermore, one can identify for this system an ``effective Planck's
constant'', $\kbar$, which is directly proportional to the period of
the laser pulsing and can therefore be adjusted to, in a sense, make
the system ``more'' or ``less'' quantum mechanical.

This so-called ``atom-optical kicked rotor'' has since been the subject
of intense investigation by a number of experimental groups in a
variety of different contexts
(see, e.g.,
\cite{Raizen1999,Ammann98,Ringot2000,dArcy2001,Williams2004}).
These investigations have in general focused on the long-time
behavior of the system; that is, on the properties of the system after
a relatively large number (at least several tens) of kicks.
For these purposes, widths of the initial momentum distributions on
the order of a few photon recoils (i.e.,
$\sigma_p\sim 4\,\hbar k_{\rm L}$, where $k_{\rm L}$ is the wave
number of the laser light) have sufficed, since the
effects under investigation have typically not been dependent on starting
from extremely precise initial states of the atomic motion. However, for detailed investigations of early-time behavior of
the kicked rotor and of certain uniquely quantum mechanical
phenomena, it is extremely desirable, or even essential,
to have yet more control over the initial state.

While most work on the kicked rotor has concentrated on differences
between classical and quantum behavior in the late-time regime early-time behavior was investigated by Shepelyansky \cite{Shepelyansky1982,Shepelyansky1987}. This work showed that
significant differences also exist in \textit{initial} diffusion
rates, and the initial quantum diffusion rate exhibits a strong
dependence on the effective Planck's constant $\kbar$.
Also this dependence can also lead to signatures in the late-time energies
\cite{Klappauf1998,dArcy2001,Williams2004}
and diffusion rates \cite{Bhattacharya2002,Daley2002R}, but for a
direct study of initial rates a very narrow initial momentum
distribution is vital from the point of view of being able
to resolve small energy changes as a function
of the system parameters after just a small number of kicks.
Furthermore, more recent theoretical work \cite{Daley2002}
has revealed that, with a very narrow initial momentum distribution,
initial diffusion rates exhibit an even richer structure (as a
function of $\kbar$) than that predicted by Shepelyansky, whose
calculations assumed broad (uniform) initial conditions.

At certain specific values of $\kbar$ -- in particular, where
$\kbar$ is a rational multiple of $4\pi$ -- quite remarkable
phenomena can occur in the form of so-called ``quantum
resonances'' and ``anti-resonances''
\cite{Izrailev1979,Izrailev1980,Casati1984,Sokolov2000,Oskay2000,Daley2002}.
These phenomena require particular initial momentum states which
evolve in such a way that during the free evolution period in
between kicks the different components of the state vector of the
system experience either identical phase shifts, or a phase shift
that alternates in sign from one momentum component to the next.
Where the phase factor is identical for all components, ballistic
energy growth is observed (quantum resonance). Where the phase
factor alternates in sign, the system returns identically to its
initial state after every second kick (quantum anti-resonance).
With a broad initial momentum distribution such resonance and
anti-resonance behavior can still be observed experimentally in
the atom-optical kicked rotor \cite{dArcy2001,Oskay2000}, but it
is far less pronounced than in the ideal case of initial momentum
eigenstates. With a dilute atomic Bose-Einstein condensate
however, it is possible to realize an initial state that is, to
all intents and purposes, a momentum eigenstate and therefore
allows a much clearer investigation of these phenomena.

In this work, we follow the suggestion of Daley and Parkins in
Ref.~\cite{Daley2002} and investigate the early-time behavior of
the atom-optical kicked rotor using a Bose-Einstein condensate to
provide a very narrow and precise initial momentum state of the
atoms. We focus on investigating the energy as a function of kick
number for specific values of effective Planck's constant, and our
results demonstrate the behaviors predicted theoretically. Note
that our work complements recent experimental studies of
atom-optical versions of (classically chaotic) nonlinear dynamical
systems which also make use of extremely narrow atomic momentum
distributions to provide very precise initial conditions
\cite{Hensinger2001,Steck2001}.

\section{The Atom Optical Kicked Rotor}

\subsection{Theoretical Model}

The basic model describing the atom-optical kicked rotor has
been described by a number of authors, and here we briefly summarize this
following the notation of Ref.~\cite{Daley2002}. A cold atomic sample
interacts with a standing wave of laser light with frequency $\omega_{\rm L}$, far-detuned from resonance. The laser is pulsed with period $T$ and pulse
profile $f(t)$. Due to the large detuning, the internal atomic dynamics can
be eliminated and the Hamiltonian determining the motion of the atoms
can be written as
\begin{equation}  \label{eqn:Ham1}
\hat{H} = \frac{\hat{p}^2}{2m} - \frac{\hbar \Omega_{\rm eff}}{8}
\cos (2k_{\rm L} \hat{x}) \sum_{n=1}^N f(t-nT),
\end{equation}
where $\hat{x}$ and $\hat{p}$ are the atomic position and momentum
operators, respectively, and $\Omega_{\rm eff}$ = $\Omega^2$/$\delta$ is
the effective potential strength,
with $\Omega /2$ the (single-beam) resonant Rabi frequency and $\delta$
the detuning from atomic resonance. We can rewrite Eq.~(\ref{eqn:Ham1}) as
a scaled dimensionless Hamiltonian in the form
\begin{equation}  \label{eqn:Ham2}
\hat{H}^{^{\prime}} = \frac{\hat{\rho}^2}{2} - k \cos (\hat\phi )
\sum_{n=1}^N f(t^\prime - n),
\end{equation}
which is standard for the kicked rotor system.
The position operator is defined by $\hat{\phi}=2k_{\rm L}$$\hat{x}$,
the momentum operator $\hat{\rho}=2k_{\rm L}$$T$$\hat{p}$/$m$, the scaled
time is $t^\prime =t/T$,
and $\hat{H}^\prime$ = (4$k_{\rm L}^{2}$$T$$^{2}$/$m$)
$\hat{H}$. The classical stochasticity parameter (or kick strength) is given
by $\kappa =\Omega_{\rm eff}$$\omega _{R}$$T$$\tau _{p}$, where $\tau _{p}$ is
the pulse length and $\omega _{R}$ = $\hbar $$k_{\rm L}^{2}$/2$m$ is the recoil
frequency. In this work $f(t^\prime )$ is taken to represent
a square pulse,
i.e. $f(t^\prime)$ = 1 for 0 $<$ $t^\prime$ $<$ $\alpha $,
where $\alpha
$ = $\tau _{p}$/$T$, in which case $k=\kappa/\alpha$. In these scaled units, we have [$\hat{\phi},\hat{\rho}$]
= $i\kbar$, with $\kbar$ = 8$\omega_{R}T$. Thus the quantum nature of the
system is reflected by an effective Planck's constant, $\kbar$, which changes
as we adjust the pulsing period $T$.

\subsection{Early-time diffusion}

In the case of the $\delta$-kicked rotor (i.e., $\alpha\rightarrow 0$,
$f(t^\prime )\rightarrow\delta (t^\prime )$) the evolution of the
system can be represented by the quantized standard
map,
\begin{eqnarray}
\label{qsmap}
\hat{\phi}_{n+1}&=&\hat{\phi}_n+\hat{\rho}_{n} \, ,\\
\hat{\rho}_{n+1}&=&\hat{\rho}_n+\kappa \sin(\hat{\phi}_{n+1}) \, ,
\end{eqnarray}
where $\hat{\phi}_n=\hat{\phi}(t^\prime =n)$
and $\hat{\rho}_n=\hat{\rho}(t^\prime =n)$,
with the values recorded immediately after the kick at $t^\prime =n$.
In this version of the standard map, the first kick occurs at
$t^\prime =1$.

In our experiment, an image of the atomic cloud allows us to determine
the momentum distribution and hence the kinetic energy after a set
number of kicks. With the change in kinetic energy between
consecutive kicks we then determine the momentum diffusion rate
\begin{equation}  \label{eqn:2kicksreal}
D(n) = \frac{\langle\hat\rho_{n}^2\rangle}{2} -
\frac{\langle\hat\rho_{n-1}^2\rangle}{2}.
\end{equation}

An analytical investigation of early-time quantum diffusion
rates in the DKR was made by Shepelyansky
\cite{Shepelyansky1982,Shepelyansky1987}, whose calculations
assumed uniform (broad) initial position and momentum distributions
and involved the evaluation of quantum correlation functions of the
form $\langle [\sin (\hat{\phi}_n),\sin (\hat{\phi}_0)]_+\rangle$ for
$n\leq 4$. From a sum of such correlation functions an estimate of
the initial quantum diffusion rate was obtained, which predicts
(broad) peaks, or resonances, as a function of $\kbar$. In particular,
prominent peaks appear in the diffusion rate where $\kbar$ is an
integer multiple of $2\pi$, together with other maxima whose number and
positions (with respect to $\kbar$) vary with kick strength $\kappa$.

More recently, Daley and Parkins \cite{Daley2002} re-examined the
early-time diffusion rates for very narrow initial momentum distributions,
as is appropriate to atom-optical experiments with Bose-Einstein condensates. They find an even more complex and interesting
structure in the diffusion rates as a function of $\kbar$,
as exemplified by their result for $D(2)$, which takes the form
\begin{eqnarray}
&& D(2) =
\frac{\kappa^2}{4} [1-J_2(K_{2q})\rme^{-2\sigma_\rho^2}
\cos(\bar{\rho}_0)]\nonumber
\\
&& -\kappa J_1(K_q)[\sigma_\rho^2 \rme^{-\sigma_\rho^2/2}
\cos(\bar{\rho}_0)+\bar{\rho}_0\rme ^{-\sigma_\rho^2/2}
\sin(\bar{\rho}_0)]\nonumber
\\
&& +\frac{\kappa^2}{2}
[J_0(K_q)-J_2(K_q)]\cos(\kbar /2) \rme^{-\sigma_\rho^2/2}
\cos(\bar{\rho}_0) , \label{eqn:diffrate2kicks}
\end{eqnarray}

where $K_q=2\kappa \sin(\kbar/2)/ \kbar$, $K_{2q}=2 \kappa
\sin(\kbar)/ \kbar$, and a Gaussian initial momentum distribution
of mean $\bar{\rho}_0$ and variance $\sigma_\rho^2$ is assumed.
The rates $D(3,4,5)$ exhibit still more structure than for $D(2)$,
but were computed numerically using wave function simulations
\cite{Daley2002R,Daley2002,Doherty2000} (which allow for finite
pulse widths and atomic spontaneous emission). Note that for a
broad initial momentum distribution $D(2)$ is independent of
$\kbar$ and given simply by $D(2)=D(1)=\kappa^2/4$.

\subsection{Quantum resonances and anti-resonances}

The phenomena of quantum resonances and anti-resonances occur for
particular values of $\kbar$, and in their ``idealized'' forms
require very specific initial conditions.
Take, for example, an initial momentum eigenstate $|\rho_0=j\kbar\rangle$,
where $j$ is an integer. Through the kicking process, this state couples
only to eigenstates $|\rho =(j+j^\prime )\kbar\rangle$, where $j^\prime$
is also an integer. If $\kbar$ is an {\em even} multiple of $2\pi$,
then it is straightforward to show that
$\exp (i\hat{\rho}^2/2\kbar )|(j+j^\prime )\kbar\rangle
=|(j+j^\prime )\kbar\rangle$,
i.e., during the free evolution period in between kicks the state
vector of the system is unchanged. This leads to ballistic
energy growth (i.e., the energy depends quadratically on kick number)
and dynamical localization does not occur.
This is known as a quantum resonance and is
related to the Talbot effect in wave optics \cite{Berry1999}. In contrast, if $\kbar$ is an {\em odd} multiple of $2\pi$, then
$\exp (i\hat{\rho}^2/2\kbar )|(j+j^\prime )\kbar\rangle
=|(j+j^\prime )\kbar\rangle$ if $j+j^\prime$ is even, and
$-|(j+j^\prime )\kbar\rangle$ if $j+j^\prime$ is odd.
In this case, the system returns identically to its initial state,
$|\rho_0=j\kbar\rangle$, after every second kick.
This effect is known as a quantum anti-resonance.

Other features (i.e., peaks or dips) appearing in the diffusion
rates as a function of $\kbar$ (see Fig.~\ref{fig:E1E2vskbar}) can
also be related to behavior such as that described above. However,
unlike quantum resonances and anti-resonances, the $\kbar$ values
for which these features occur depend on the kick strength $\kappa$
in a nontrivial manner \cite{Daley2002}.

%\newpage

\section{Experiment}

We focus on the behavior of the energy as a function of kick number
and of the effective Planck's constant, $\kbar$.
The kick strength is fixed by keeping the laser intensity constant and varying
both the pulsing period $T$ (which is proportional to the effective
Planck's constant) and pulse length $\tau _{p}$, such that the product
$T\tau_p$ is a constant.
The experiment is
performed with a Bose condensate of approximately ${10^{4}}$
${}^{87}$Rb atoms in the $F$~=~2, ${m_F}$~=~2 hyperfine state. The
condensate is formed (as described in Ref.~\cite{Martin1999}, but
with minor modifications \cite{Mellish2003}) in a time-averaged
orbiting potential trap with harmonic oscillation frequencies of
$\omega _{r}$/{2$\pi $}~=~71~Hz radially and $\omega _{z}$/{2$\pi
$} ~=~201~Hz axially. After radio frequency evaporation to form a
Bose condensate, the trap is relaxed to $\omega _{r}$/{2$\pi
$}~=~32~Hz and $\omega _{z}$/{2$\pi $}~=~91~Hz over a period of
200~ms.

\subsection{Kicking}

Once formed, the condensate is released from the trap and exposed
to pulsed optical standing waves after 1.7~ms of free expansion
(so that condensate mean-field effects can essentially be ignored
\cite{Stenger1999}).  For our parameters, the momentum FWHM of the
Bose condensate is 0.03$\times 2\hbar k_{\rm L}$. These standing
waves are generated by two counterpropagating laser beams with
parallel linear polarizations, derived from a single beam which is
detuned 1.48~GHz from the 5S$_{1/2}$, $F$~=~2 $\rightarrow$
$5{P}_{3/2}$, ${F}$$^{\prime }$~=~3 transition. For a chosen
$\kappa$ value, the pulse period is scanned from 21.12~$\mu$s
($\kbar =4$) to the quantum resonance at ${T}$ = 66.38~$\mu$s
($\kbar =4\pi$). Consequently the pulse length is varied from
1.25~$\mu $s to 400~ ns. There are limitations on the precise
values of $T$ and $\tau_p$ caused by the incremental stepsize
values of the pulse generator. The laser detuning and intensity
were chosen to give the desired kicking strength while maintaining
a negligible spontaneous emission rate ($<$ 34 s$^{-1}$). The
momentum distribution is determined using time-of-flight (4~ms of
free expansion) and absorption imaging of the atomic sample. When
the condensate is released from the magnetic trap it receives an
(unwanted) impulse corresponding approximately to 12.5$\pm$0.5
mm/s, as determined by Bragg scattering \cite{Geursen2003}. In
order to apply a standing wave which is stationary with respect to
the condensate the frequency difference between the two laser
beams is adjusted to remove the relative motion. A double-pass
acoustic-optic modulator is used in each beam for altering its
frequency and switching the optical potential on and off.

\subsection{Energy measurements}

\begin{figure}[htb]
\begin{center}
\psfrag{a }{\raisebox{-1mm}{\tiny$0\,\hbar k_L$}} \psfrag{b
}[-30cm][l]{\raisebox{3mm}{\tiny$-2\,\hbar k_L$}} \psfrag{c
}[-30cm][l]{\raisebox{-1mm}{\tiny$+2\,\hbar k_L$}} \psfrag{d
}[-30cm][l]{\raisebox{3mm}{\tiny$-4\,\hbar k_L$}} \psfrag{e
}[-30cm][l]{\raisebox{3mm}{\tiny$-4\,\hbar k_L$}}
\psfrag{w}[-50cm][l]{\tiny 1 kick} \psfrag{x}[-50cm][l]{\tiny 2
kicks} \psfrag{y}[-50cm][l]{\tiny 3 kicks}
\psfrag{z}[-50cm][l]{\tiny 4 kicks} \psfrag{Momentum
States}{\hspace{-7mm}\tiny Momentum States}
\psfrag{Atomic}{\hspace{-10mm}\tiny Atomic Population (arb units)}

\begin{minipage}{0.49\linewidth}
\center
\includegraphics[width=0.6\linewidth,height=0.7\linewidth]{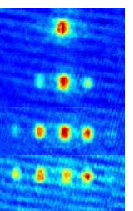}
\end{minipage}
\begin{minipage}{0.49\linewidth}
\hspace{15mm}
\includegraphics[width=1.0\linewidth]{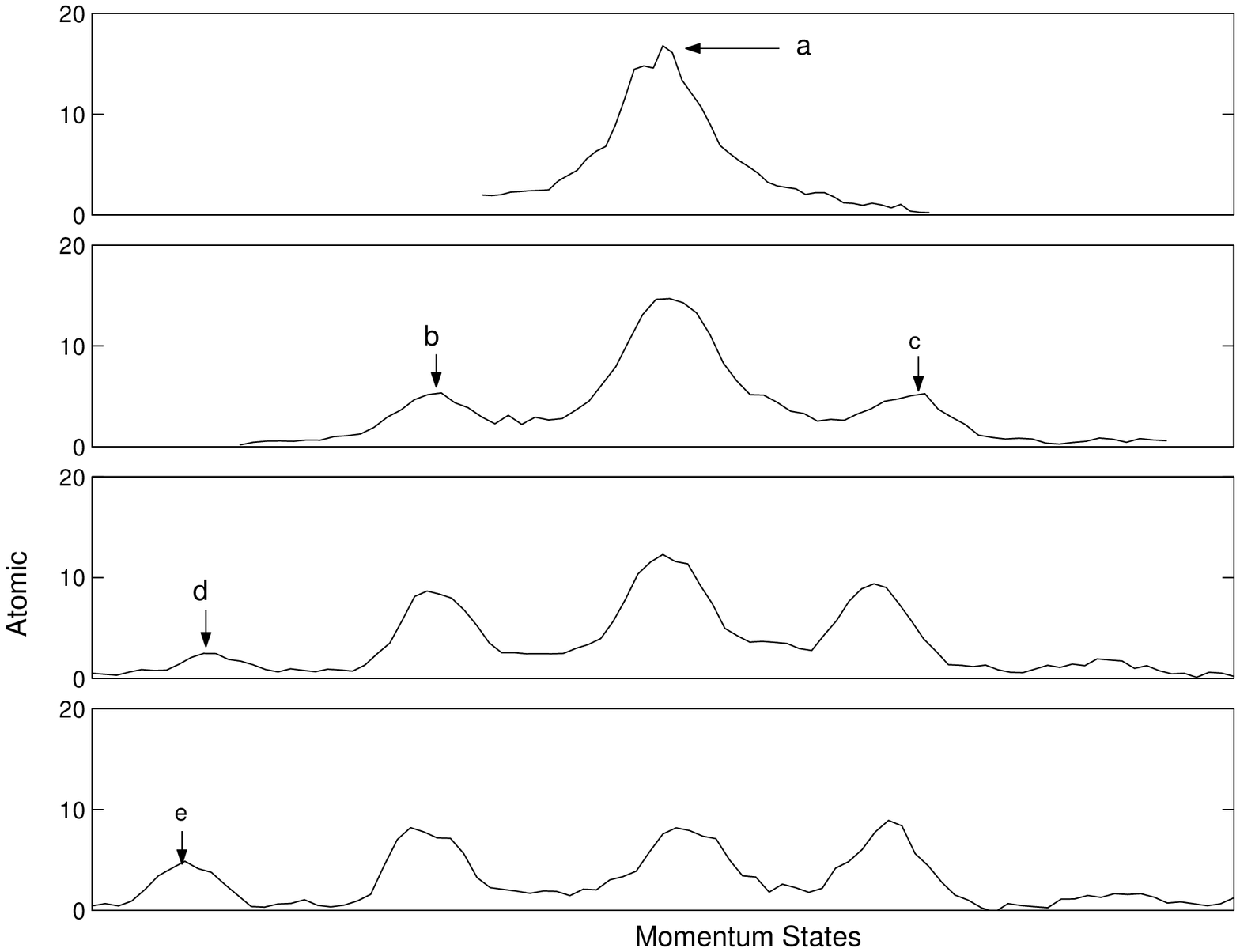}
\end{minipage}
\end{center}
\caption{(a) Typical time-of-flight images of atomic clouds seperated in momentum by $2\hbar k_{L}$ for 1(top) to 4(bottom) kicks (b) Corresponding momentum distributions. This particular case illustrates enhanced energy growth close to the quantum resonance at $\kbar = 4\pi$.}
\label{fig:ExptDataQR}
\begin{picture}(0,0)
\put(-65,70){\scriptsize(a)}
\put(60,70){\scriptsize(b)}
\end{picture}
\end{figure}

The kicks given to the atomic sample populate
momentum classes separated by $2\hbar k_{\rm L}$ (as shown in Fig.~\ref{fig:ExptDataQR}).
Following the kicks, we determined the kinetic energy of the atomic sample,
which entailed counting the number of atoms in each momentum
state and multiplying by the energy of that state. Subsequently, to obtain the
average kinetic energy per atom this
energy value was divided by $(2\hbar k_{\rm L})^{2}$ and the total number
of atoms, and then multiplied by $m\kbar^{2}$ (where $m$ is the
mass of the rubidium atom).
This gave us an overall average energy in dimensionless units,
$E={\langle\rho^{2}\rangle}/{2}$.

\section{Experimental Results}
\begin{figure}[htp]
\begin{center}
\psfrag{1st kick}{\tiny 1 kick} \psfrag{2nd kick}{\tiny 2 kicks}
\psfrag{Energy}{\hspace{-3mm}\tiny Energy <$\rho^2$>/2}
\psfrag{Effective}{\hspace{-5mm}\tiny Effective Planck's Constant}

\epsfig{file=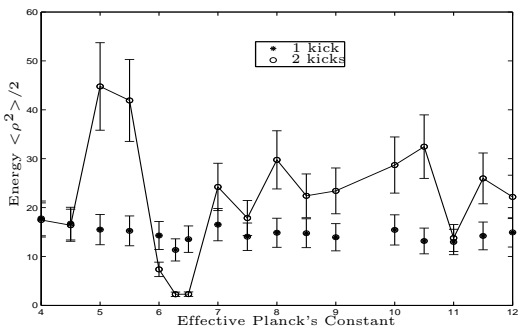,width=0.8\linewidth,height=0.5\linewidth}
\end{center}
\caption{ Experimental energies (in dimensionless units) after 1
and 2 kicks, $E(1)$ and $E(2)$, at the lower kicking strength.
After 1 kick the energies are approximately constant as a function
of the effective Planck's constant $\kbar$ (theory predicts
$E(1)=\kappa^2/4$, a constant). For a broad initial momentum
distribution, theory predicts simply that $E(2)=2E(1)=\kappa^2/2$,
but for a very narrow initial momentum distribution the energy
$E(2)$ is heavily dependent on $\kbar$ and exhibits pronounced
peaks and dips. The error bars reflect shot-to-shot variation in
the atom number and fluctuation in the laser intensity.}
\label{fig:E1E2vskbar}
\end{figure}

We begin by highlighting the rich initial diffusion behavior
of our kicked rotor system with experimental data for the energy after
one and two kicks, $E(1)$ and $E(2)$, as a function of the effective
Planck's constant, $\kbar$, for a fixed value of the kicking strength
$\kappa$. This data, plotted in Fig.~\ref{fig:E1E2vskbar},
confirms the prediction Ref.~\cite{Daley2002} of a uniform value
for $D(1)=E(1)-E(0)$, but a strong dependence of $D(2)=E(2)-E(1)$ on
$\kbar$, given a sufficiently narrow initial momentum distribution.
Similar behavior is obtained for the higher kicking strength, as we
shall see later.

The energy after one kick is given theoretically by
$E(1)=\kappa^2/4$, which allows us to infer a value $\kappa
=7.7\pm 0.6$ ($11.7\pm 2$) from the experimental data for the
lower (higher) kicking strength. These values are consistent with
those calculated from the laser intensity, detuning and pulse
details.
Note that for the one-kick data the duration of the kicking pulse was
chosen for each $\kbar$ value to match that used for sequences of two
or more kicks, so the one-kick energies do in fact correspond to
different experimental conditions. However, in dimensionless units the
energy after one kick ($\langle\rho_1^2\rangle /2$) is predicted to
be constant as a function of $\kbar$.

Before continuing, we note that a related experiment in which a released Bose-Einstein condensate of sodium atoms was subjected to a sequence of two
standing-wave laser pulses separated by a varying time delay was recently reported by Deng {\em et al}. \cite{Deng1999}.
While effects related to those displayed above could be inferred
from their results, their work was not set in the context
of the quantum kicked rotor -- if one does so, then the experiment
they performed corresponds to varying $\kbar$ and $\kappa$
simultaneously (since both parameters are proportional to the
kicking period $T$, and the laser intensity and $\tau_p$ were fixed
for their measurements).

\subsection{Energy versus kick number}

\begin{figure}[htp]
%\psfrag{x}[-50cm][l]{\tiny 2nd kick}
\psfrag{Energy}{\hspace{-5mm}\tiny Energy <$\rho^2$>/2}
\psfrag{Kick Number}{\hspace{-5mm}\tiny Kick Number}
\psfrag{a}{\tiny a}
\psfrag{b}{\tiny b}
\psfrag{c}{\tiny c}

\begin{center}
\epsfig{file=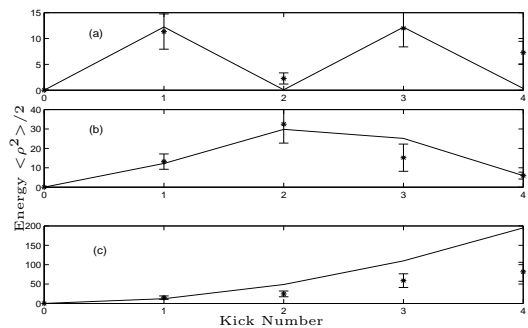,width=0.8\linewidth,height=0.5\linewidth}
\end{center}
\caption{Energy versus kick number for the lower kicking strength
$\kappa$ =7.7 (a) $\kbar =6.211$, (b) $\kbar =10.523$, and (c)
$\kbar =12.573$. Crosses are experimental data, while the solid
lines are the results of numerical simulations with $\kappa =7$.
Plot (a) illustrates anti-resonance behavior at $\kbar\simeq
2\pi$, while plot (b) demonstrates the presence of a period-4
anti-resonance close to $\kbar =10.5$. Plot (c) illustrates
enhanced energy growth close to the quantum resonance at
$\kbar=4\pi$.} \label{fig:EvsN15}
\end{figure}

\begin{figure}[htp]
\psfrag{Energy}{\hspace{-5mm}\tiny Energy <$\rho^2$>/2}
\psfrag{Kick Number}{\hspace{-5mm}\tiny Kick Number}
\begin{center}
\epsfig{file=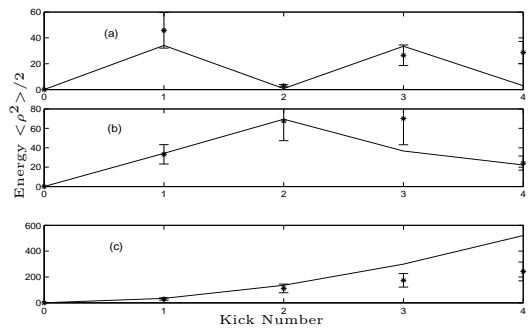,width=0.8\linewidth,height=0.5\linewidth}
\end{center}
\caption{
Energy versus kick number for the higher
kicking strength, $\kappa =11.7$ (a) $\kbar =6.211$, (b) $\kbar =10.992$, and
(c) $\kbar =12.573$. Similar resonance type behavior is observed in figure~\ref{fig:EvsN15} as expected.}
\label{fig:EvsN25}
\end{figure}

The energy as a function of kick number exhibits a complex variety
of behaviors as the effective Planck's constant is varied (for
fixed $\kappa$). Examples are plotted in Figs.~\ref{fig:EvsN15}
and \ref{fig:EvsN25}, where experimental energies for up to four
kicks are shown. Close to $\kbar =2\pi$ (Figs.~\ref{fig:EvsN15}(a)
and \ref{fig:EvsN25}(a)) one observes the anti-resonance
phenomenon described earlier, in which the system returns
(approximately) to its initial state after every second kick --
this manifests itself as an oscillation in the energy. In
contrast, near $\kbar=4\pi$ (Figs.~\ref{fig:EvsN15}(c) and
\ref{fig:EvsN25}(c)) one sees a continual growth in the energy as
a consequence of the phenomenon of quantum resonance. For the
lower kicking strength, we also observe behavior suggesting a
period-4 anti-resonance close to $\kbar=10.5$
(Fig.~\ref{fig:EvsN15}(b)). Numerical simulations confirm this
behavior, for $\kappa=7$, at a value of $\kbar$ in this vicinity.
For the larger kicking strength similar behavior occurs to a
certain extent, but is most pronounced at slightly larger values
of $\kbar$. Unfortunately, falling signal-to-noise and the effects
of a finite initial mean momentum means that it is very difficult
to extract useful energy values beyond about 4 kicks and hence to
rigorously confirm this higher-order behavior (at least with the
experimental setup that was used).

\begin{figure}[htb]
\psfrag{Energy}{\hspace{-5mm}\tiny Energy <$\rho^2$>/2}
\psfrag{Kick Number}{\hspace{-5mm}\tiny Kick Number}
\begin{center}
\epsfig{file=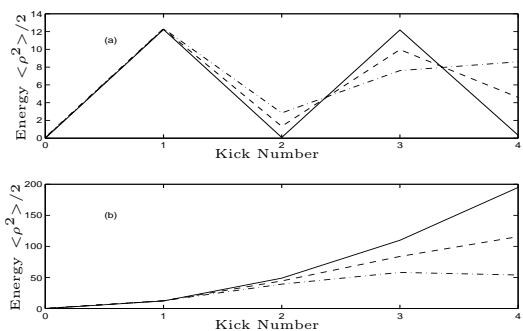,width=0.8\linewidth,height=0.5\linewidth}
\end{center}
\caption{
Numerical simulations of energy as a function of kick number for (a) $\kbar =6.211$ and (b) $\kbar =12.573$,
with $\kappa =7$ and $\bar{p}_0=0$ (solid), $\bar{p}_0=0.1\,\hbar k_{\rm L}$
(dashed), and $\bar{p}_0=0.15\,\hbar k_{\rm L}$ (dot-dashed).
The quantum anti-resonance and resonance phenomena are clearly very sensitive
to the initial mean momentum of the atomic cloud.
}
\label{fig:finitep0}
\end{figure}

Also plotted in Figs.~\ref{fig:EvsN15} and \ref{fig:EvsN25} are
the results of numerical simulations for the appropriate values of
$\kbar$ and $\alpha$, and for $\kappa =7.0$ and $\kappa =11.7$,
respectively. Quantitative agreement between theory and experiment
is reasonably good, but notable deviations appear after several
kicks for the anti-resonance and resonance at $\kbar\simeq 2\pi$
and $4\pi$, respectively. We believe that these deviations are due
partly to fluctuations in the initial mean momentum of the atomic
cloud relative to the laser standing wave. Inspection of the
images of the kicked atomic clouds reveals that the position of
the ``zero momentum'' peak can fluctuate from shot-to-shot by an
amount of up to a few tenths of an atomic recoil. This was
confirmed by Bragg scattering measurements \cite{Geursen2003} and
is associated with the precise details of the switching-off of the
time-averaged orbiting potential trap used to confine the initial
Bose-Einstein condensate. As shown in \cite{Daley2002}, the
quantum anti-resonance and resonance phenomena are particularly
sensitive to any non-zero initial mean momentum of the atomic
ensemble. In Fig.~\ref{fig:finitep0} we demonstrate this
sensitivity with simulations for the parameters associated with
Figs.~\ref{fig:EvsN15}(a) and \ref{fig:EvsN15}(c), but now for
initial mean momenta $\bar{p}_0=0.1\,\hbar k_{\rm L}$ and
$\bar{p}_0=0.15\,\hbar k_{\rm L}$. The width of the initial
momentum distribution is the same as in the experiment. As shown,
with a finite value of $\bar{p}_0$, one finds behavior more
consistent with the experimental data.

\begin{figure}[htb]
\psfrag{E(2)}{\hspace{-5mm}\tiny E(2)=<$\rho^{2}_{2}$>/2}
\psfrag{Effective}{\hspace{-5mm}\tiny Effective Planck's Constant}
\begin{center}
\epsfig{file=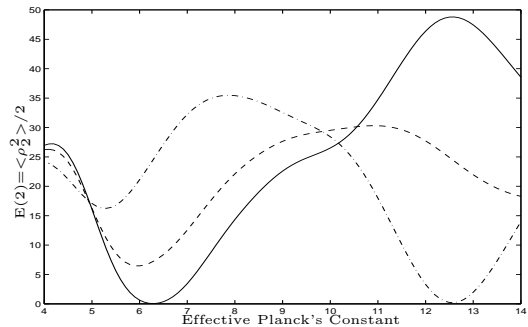,width=0.8\linewidth,height=0.5\linewidth}
\end{center}
\caption{
Theoretical predictions using Eq.~(\ref{eqn:diffrate2kicks}) for the energy after 2 kicks, $E(2)$, versus effective
Planck's constant $\kbar$, for $\kappa =7$ and $\bar{p}_0=0$ (solid),
$\bar{p}_0=0.25\,\hbar k_{\rm L}$ (dashed), and
$\bar{p}_0=0.5\,\hbar k_{\rm L}$
(dot-dashed). These results further illustrate the
sensitivity of the energy growth rate to the initial mean momentum of
the atomic cloud, particularly at the quantum resonance ($\kbar =4\pi$).
}
\label{fig:E2theory_p0}
\end{figure}

The sensitivity of the energy to the initial mean momentum of the
atomic cloud adds an extra variable to the problem, over which we
at present have little control. To further emphasize the dependence of the
energy on the initial mean momentum, in Fig.~\ref{fig:E2theory_p0}
we plot the energy after two kicks, $E(2)$, using the theoretical result
of Eq.~(\ref{eqn:diffrate2kicks}) for several values of the initial mean
momentum, $\bar{p}_0$.
This plot highlights the extreme sensitivity to initial motion of the
quantum resonance at $\kbar =4\pi$, which in fact changes to an {\em
anti-resonance} when $\bar{p}_0=0.5\,\hbar k_{\rm L}$.
It follows that, in order to see resonance and antiresonance behavior
controllably in their clearest forms, one requires very precise control
over the initial mean motion of the atomic cloud.

\section{Conclusion}
In summary our results show a rich structure of resonances as a function of the effective
Planck's constant and kick number for the atom-optical kicked rotor with a narrow initial
momentum distribution. For two kicks we have shown that the resonant behavior is a more complex
than those predicted by Shepelyansky for a system with a broad initial momentum distribution.
Moreover we observed quantum anti-resonance ($\kbar = 2\pi$) and quantum resonance ($\kbar = 4\pi$)
 in the energy as a function of $\kbar$ for different kick strengths. Comparison between theory and
  experiment showed reasonable agreement, but by introducing an initial mean momentum to the
  numerical model better agreement could be obtained. Consequently, we have shown that the quantum
  features are very sensitive to precise initial conditions. One possible method of eliminating
  the initial mean momentum of the condensate would be to perform the measurements with the
  condensate still confined by the magnetic trap. One difficulty with this is that for our
  apparatus, condensate micro-motion will be introduced \cite{Geursen2003}. In addition, mean
  field effects will occur, although the nonlinearity associated with these provide an opportunity
   to investigate instability of the condensate wave function \cite{Niu2003}.

\begin{acknowledgments}
The authors thank A. Daley for help with initial calculations. We acknowledge the support of the Marsden Fund of the Royal
Society of New Zealand through grants 02UOO910 and UOA016.
\end{acknowledgments}

\end{document}